\documentclass[aps,prb,showpacs,floatfix,twocolumn,byrevtex,superscriptaddress]{revtex4-1}
\usepackage{amsmath}
\usepackage{amssymb}
\usepackage{amstext}
\usepackage{amsopn}
\usepackage{amsfonts}
\usepackage{amsxtra}
\usepackage[english]{babel}
\usepackage{amsmath}
\usepackage{graphicx}
\usepackage{float}
\usepackage{bm}
\usepackage{multirow}
\usepackage{dcolumn}
\usepackage{hyperref}
\usepackage{CJK}

\begin{document}
\title{Observation of topological nodal-line semimetal in YbMnSb$_{2}$ through optical spectroscopy}
\author{Ziyang Qiu}
\affiliation{Beijing National Laboratory for Condensed Matter Physics, Institute of Physics, Chinese Academy of Sciences, P.O. Box 603, Beijing 100190, China}
\affiliation{School of Physical Sciences, University of Chinese Academy of Sciences, Beijing 100049, China}
\author{Congcong Le}
\affiliation{Beijing National Laboratory for Condensed Matter Physics, Institute of Physics, Chinese Academy of Sciences, P.O. Box 603, Beijing 100190, China}
\affiliation{Kavli Institute of Theoretical Sciences, University of Chinese Academy of Sciences, Beijing, 100049, China}
\author{Zhiyu Liao}
\affiliation{Beijing National Laboratory for Condensed Matter Physics, Institute of Physics, Chinese Academy of Sciences, P.O. Box 603, Beijing 100190, China}
\affiliation{School of Physical Sciences, University of Chinese Academy of Sciences, Beijing 100049, China}
\author{Bing Xu}
\affiliation{Beijing National Laboratory for Condensed Matter Physics, Institute of Physics, Chinese Academy of Sciences, P.O. Box 603, Beijing 100190, China}
\author{Run Yang}
\affiliation{Beijing National Laboratory for Condensed Matter Physics, Institute of Physics, Chinese Academy of Sciences, P.O. Box 603, Beijing 100190, China}
\affiliation{School of Physical Sciences, University of Chinese Academy of Sciences, Beijing 100049, China}
\author{Jiangping Hu}
\affiliation{Beijing National Laboratory for Condensed Matter Physics, Institute of Physics, Chinese Academy of Sciences, P.O. Box 603, Beijing 100190, China}
\affiliation{Kavli Institute of Theoretical Sciences, University of Chinese Academy of Sciences, Beijing, 100049, China}
\affiliation{Songshan Lake Materials Laboratory, Dongguan, 523808, China}
\author{Yaomin Dai}
\affiliation{Center for Superconducting Physics and Materials, National Laboratory of Solid State Microstructures and Department of Physics, Nanjing University, Nanjing 210093, China}
\author{Xianggang Qiu}
\email[]{xgqiu@iphy.ac.cn}
\affiliation{Beijing National Laboratory for Condensed Matter Physics, Institute of Physics, Chinese Academy of Sciences, P.O. Box 603, Beijing 100190, China}
\affiliation{School of Physical Sciences, University of Chinese Academy of Sciences, Beijing 100049, China}
\affiliation{Songshan Lake Materials Laboratory, Dongguan, 523808, China}
%

\begin{abstract}
The optical properties of YbMnSb$_{2}$ have been measured in a broad frequency range from room temperature down to 7~K. With decreasing temperature, a flat region develops in the optical conductivity spectra at about 300~cm$^{-1}$, which can not be described by the well-known Drude-Lorentz model. A frequency-independent component has to be introduced to model the measured optical conductivity. Our first-principles calculations show that YbMnSb$_{2}$ possesses a Dirac nodal line near the Fermi level. A comparison between the measured optical properties and calculated electronic band structures suggests that the frequency-independent optical conductivity component arises from interband transitions near the Dirac nodal line, thus demonstrating that YbMnSb$_{2}$ is a Dirac nodal-line semimetal.
\end{abstract}


\maketitle

%

\section{Introduction}
In the past few years, the research of topological materials, such as topological insulators~\cite{Hasan2010,Bansil2016,Qi2011}, Dirac semimetals~\cite{Armitage2018,Xu2018,Raza2019}, Weyl semimetals~\cite{Armitage2018,Weng2015,Burkov2011-2} as well as nodal-line semimetals~\cite{Burkov2011}, has inspired a great deal of interests in condensed matter physics, due to not only the fascinating physics they exhibit, but also their potential applications in electronics and quantum computing. Topological materials exhibit many unusual properties. For example, electrons on the surface of a topological insulator suffer no back scattering due to spin-momentum locking~\cite{Konig2007,Hsieh2009,Hor2010}; the surface state of a Weyl semimetal features Fermi arcs that connect the bulk Weyl fermions with opposite chiralities~\cite{Weng2015,HuangXC2015}; large negative megnetoresistance arises from the Adler-Bell-Jackiw chiral anomaly in Weyl semimetals~\cite{Ezawa2017,Lv2017,Aji2012}. Among topological materials, one important class is the nodal-line semimetals~\cite{Burkov2011,Fang2015,Armitage2018,Weng2016}. Because of the extension of quasi-two dimensional Dirac bands along lines in Brillouin zone, nodal-line semimetals can be considered as precursors of many other topological phases, such as topological insulators and Weyl semimetals~\cite{Weng2016,Armitage2018}. Even though there are various predicted nodal-line materials (such as Cu$_{3}$PdN~\cite{Yu2015}, SrIrO$_{3}$~\cite{Chen2016}, CaAgP, and CaAgAs~\cite{Wang2017,XuN2018}), the experimental evidences have only been obtained in ZrSiS~\cite{Schilling2017-2,Topp2017,ChenC2017}, PbTaSe$_{2}$~\cite{Bian2016} and NbAs$_{2}$~\cite{Shao2019}, and more efforts are needed to search other nodal-line semimetals.


Recently, the $A$Mn$P_{2}$ ($A$=Ca, Sr, Ba, Eu and Yb, $P$=Bi and Sb) family with highly anisotropic Dirac dispersion has attracted much attention~\cite{Park2011,WangK2012,Li2016,Wang2016,Wang2011,May2014,Chinotti2016,Lee2013,Liu2016,Chaudhuri2017,Qiu2018,Kealhofer2018,Wang2018}. Due to the magnetic order induced by the Mn lattice~\cite{Guo2014,Wan2011,Lee2013}, these compounds have been intensively studied with the aim of realizing more topological quantum phases arising from broken time reversal symmetry, such as the magnetic Weyl fermions in YbMnBi$_{2}$~\cite{Borisenko2015,Chinotti2016}. However, a recent optical study revealed signatures of a gapped Dirac dispersion in this material, in conflict with what is expected for a Weyl semimetal~\cite{Chaudhuri2017}.

In contrast to YbMnBi$_{2}$, the nitrogen sister compound YbMnSb$_{2}$ has relatively weaker spin-orbital coupling (SOC) effect from Sb atoms, which is more likely to host the massless Dirac or Weyl fermions. The nontrivial properties have been observed through Hall effect, magnetotransport measurements, as well as angle resolved photoemission spectroscopy (ARPES)~\cite{Kealhofer2018,Wang2018}. We have successfully synthesized the single crystal YbMnSb$_{2}$ and carefully examined its optical conductivity using infrared spectroscopy, which is a powerful tool to study the excitations near the Fermi level in topological materials~\cite{Carbotte2017,Hosur2012,Ashby2014,Armitage2018}. Several features in both the reflectivity and optical conductivity spectra are recognized at low frequency upon cooling. The emergence of a peculiar constant background optical conductivity has been observed, and we argue that it is associated with the nodal-line Dirac dispersion in YbMnSb$_{2}$. In combination with theoretical calculations, we conclude that YbMnSb$_{2}$ is a robust Dirac nodal-line semimetal.

\section{experiment}
High-quality single crystals of YbMnSb$_{2}$ with good cleavage $ab$ planes were grown by Sb self-flux method~\cite{Wang2018}. The dimension of the obtained single crystals is about 4 mm $\times$ 4 mm $\times$ 0.05 mm. Resistivity measurement was carried out on a Quantum Design physical property measurement system(PPMS). The crystal structure was characterized by x-ray diffraction (XRD) using a PANalytical diffractometer with Cu $K_{\alpha}$ radiation at room temperature.

The frequency-dependent reflectivity from a freshly cleaved surface has been measured at a near-normal angle of incidence on a Bruker 80v Fourier transform infrared (FTIR) spectrometer. By using an \emph{in situ} gold evaporation technique~\cite{Homes1993}, data from 50 to 15\,000 cm$^{-1}$ were collected at 14 different temperatures from 7 to 295~K. The visible-UV range (10\,000-30\,000 cm$^{-1}$) reflectivity was measured at room temperature with an Avaspec 2048 $\times$ 14 optical fiber spectrometer. We obtain the real part of the optical conductivity $\sigma_{1}(\omega)$ through a Kramers-Kronig analysis of $R(\omega)$. Since the scattering rate of the Drude component is about 20~cm$^{-1}$, much smaller than $\nu_{min} \thickapprox$ 100 cm$^{-1}$, a set of Lorentz oscillators instead of the commonly Hagen-Rubens form ($R=1-A\sqrt{\omega}$) are applied to extrapolate the measured low-frequency reflectivity~\cite{Chaudhuri2017,Schilling2017}. Above the highest-measured frequency (30\,000~cm$^{-1}$), $R(\omega)$ is assumed to be constant up to 12.4~eV, above which a free-electron response ($\omega^{-4}$) was used.

\section{results}
\subsection{Single crystals and resistivity}
\begin{figure}
  \centering
  \includegraphics[width=0.9\columnwidth]{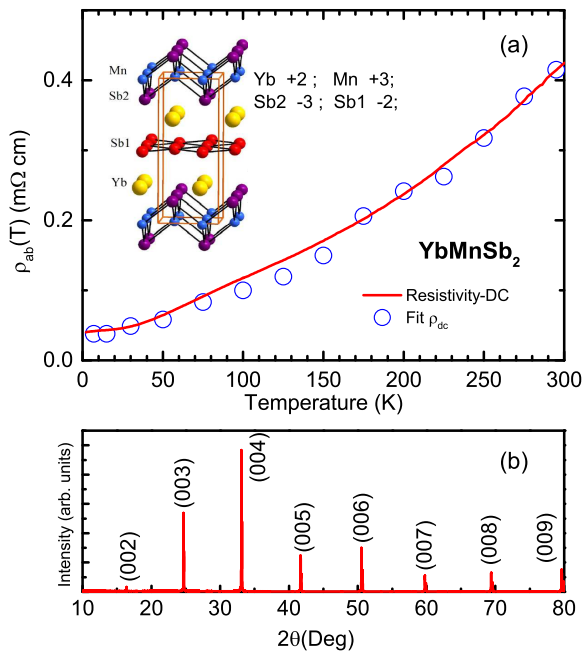}\\
  \caption{(a) Temperature dependence of resistivity of YbMnSb$_{2}$, the blue circles denote the zero-frequency extrapolation of its optical conductivity. The inset shows the schematic crystal structure of YbMnSb$_{2}$. (b)Experimental XRD pattern of a flat YbMnSb$_{2}$ crystal suggests it high quality with good cleavage $ab$ plane.}
  \label{resistivity}
\end{figure}

Figure~\ref{resistivity}(a) shows the measured temperature-dependent resistivity $R(T)$ of YbMnSb$_{2}$. $R(T)$ decreases upon cooling, indicating the metallic behavior. The crystal structure of YbMnSb$_{2}$ has been indexed to the $P4/nmm$ space group~\cite{Wang2018,Kealhofer2018}. The inset of Fig.~\ref{resistivity}(a) plots the crystal structure with alternately stacked YbSb and MnSb layers. The single-crystal XRD pattern in Fig.~\ref{resistivity}(b) displays the good $(00l)$ reflections. The sharp and clean XRD pattern suggests our single crystals of YbMnSb$_{2}$ are of high quality with $ab$ plane~\cite{Wang2018}.

\subsection{Reflectivity and optical conductivity}
\begin{figure}
  \centering
  \includegraphics[width=0.9\columnwidth]{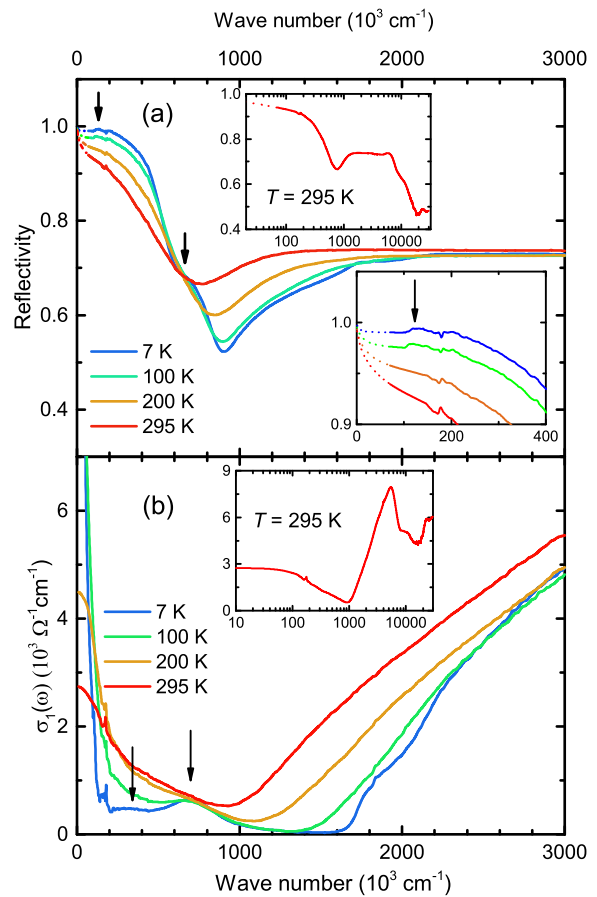}\\
  \caption{ (Color online) Reflectivity (a) and the real part $\sigma_{1}(\omega)$ of the optical conductivity (b) of YbMnSb$_{2}$ at four representative temperatures. The upper inset of (a) and the lower inset of (b) show the reflectivity and the optical conductivity over a broad frequency range at 295~K, respectively. The lower inset of (a) depicts an enlarged view of the low frequency features. The color dashed lines are the sets of Lorentz oscillators for extrapolation.}
  \label{reflectivity}
\end{figure}

The measured in plane frequency-dependent reflectivity up to 3\,000~cm$^{-1}$ at four selected temperatures is displayed in Fig.~\ref{reflectivity}(a). The high reflectivity at low frequency that approaches unity below 200~cm$^{-1}$ at low temperature is a signature of metallic behavior. $R(\omega)$ drops sharply at about 400~cm$^{-1}$ with a narrow plasma edge-like feature pointing to a  low scattering rate. In addition, the plasma edge exhibits a strong temperature dependence~\cite{Xi2014,Xu2018}: it increases from 700~cm$^{-1}$  at 295~K to 900~cm$^{-1}$ upon cooling to 7~K. At low temperature, besides the usual Drude response, two abnormal features are identified in the spectrum at about 100 and 600~cm$^{-1}$ (denoted by the arrows).

Figure~\ref{reflectivity}(b) depicts $\sigma_{1}(\omega)$ below 3\,000~cm$^{-1}$. At low frequency ($\leq$ 1\,500~cm$^{-1}$), a sharp Drude-like response is observed in $\sigma_{1}(\omega)$. Above 1\,500~cm$^{-1}$, the increase of $\sigma_{1}(\omega)$ indicates the emergence of interband transitions. As the temperature is reduced, the Drude response narrows, and two distinct features emerge: a flat optical conductivity from 200 to 500~cm$^{-1}$ and a Lorentz-like response at about 700~cm$^{-1}$ (arrowed in Fig.~\ref{reflectivity}(b)). The detailed discussion on the possible origins of these features will be given below.

\subsection{Low frequency response}

\begin{figure}[tb]
  \centering
  \includegraphics[width=1\columnwidth]{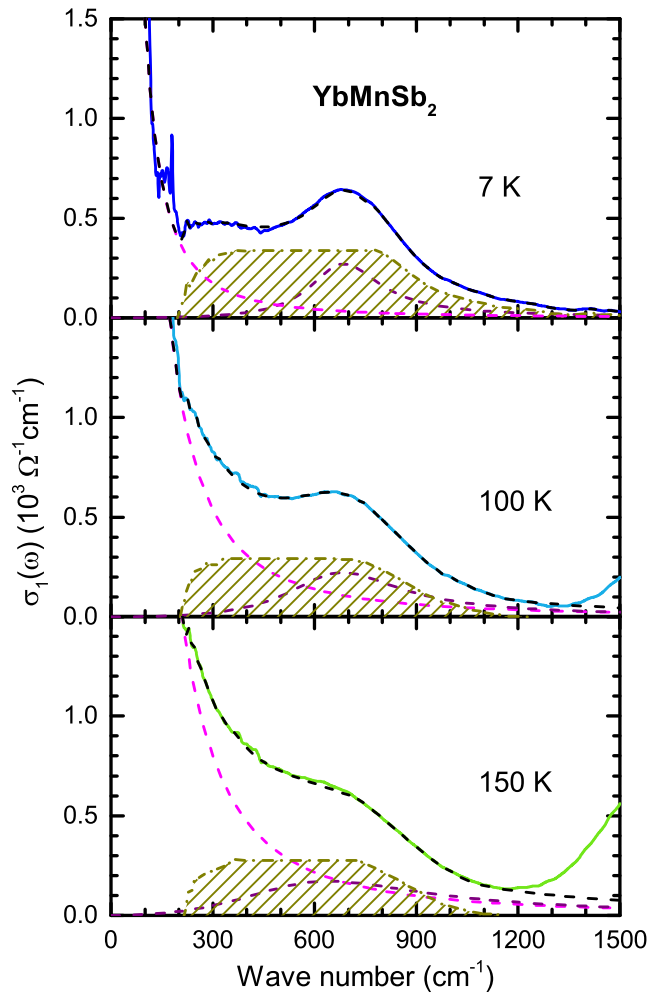}
  \caption{Examples of fit (black dashed curve) to $\sigma_{1}(\omega)$ (blue solid curve) in the frequency range of 0-1\,500~cm$^{-1}$, which is decomposed into one Drude (magenta dashed curve), one Lorentzian (purple dashed curve), and a constant term (dark yellow dash dot curve). The infrared-active phonons at around 150~cm$^{-1}$ are ignored during our fitting.}
  \label{fit}
\end{figure}

Usually, a Drude-Lorentz model~\cite{Dressel2002} may be employed to quantitatively analyze the low-frequency $\sigma_{1}(\omega)$ (0-1\,500~cm$^{-1}$) of YbMnSb$_{2}$,
\begin{equation}
\label{Drude-Lorentz}
  \sigma_{1}(\omega)=\frac{2\pi}{Z_{0}}[\sum_{k}\frac{\Omega_{p,k}^{2}}{\tau_{k}(\omega^{2}+\tau_{k}^{-2})}+
  \sum_{j}\frac{\gamma_{j}\omega^{2}\Omega_{j}^{2}}{(\omega_{j}^{2}-\omega^{2})^{2}+\gamma_{j}^{2}\omega^{2}}],
\end{equation}
where Z$_{0}$=377 $\Omega$ is the vacuum impedance. The first term describes a sum of free-carrier Drude responses, where $\Omega_{p}=\sqrt{4\pi ne^{2}/m^{*}}$ is the plasma frequency($n$ is a carrier density and m$^{*}$ is an effective mass), and $1/\tau$ is the scattering rate of carriers. The second term corresponds to a sum of Lorentz oscillators, with $\omega_{j}$, $\gamma_{j}$ and $\Omega_{j}$ being the resonance frequency, width and strength of the $j$th vibration or bound excitation.

We have attempted to fit the measured $\sigma_{1}(\omega)$ with one Drude and two Lorentz terms. However, this Drude-Lorentz model fails in describing the measured $\sigma_{1}(\omega)$, particularly in the range of 200-1\,500~cm$^{-1}$. We noticed that $\sigma_{1}(\omega)$ exhibits a constant background in the 200-500~cm$^{-1}$ range. Therefore, a constant component has to be introduced into the Drude-Lorentz model to fit $\sigma_{1}(\omega)$ at all measured temperatures. The solid circles in Fig.~\ref{resistivity}(a) represent the zero-frequency value of the fit. Compared to the dc resistivity from transport measurements (the solid curve), the good agreement between the optical and transport results implies the modeling is reliable.

\begin{figure}[tb]
  \centering
  \includegraphics[width=1\columnwidth]{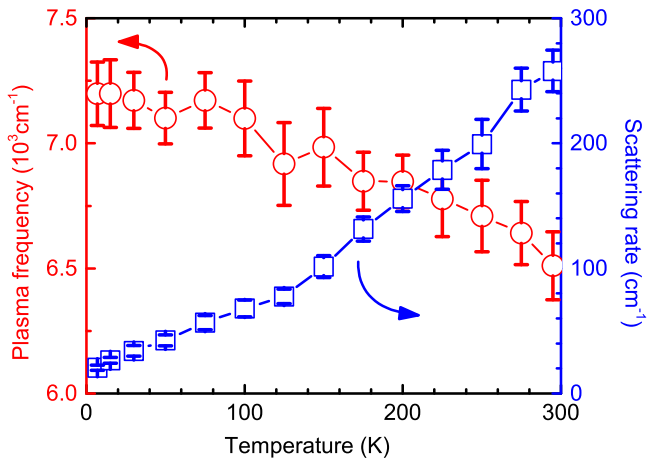}\\
  \caption{ (Color online) The corresponding scattering rate and plasma frequency of the Drude component at various temperatures extracted from the model fitting.}
  \label{fit data}
\end{figure}

Figure~\ref{fit} displays the fitting results at three representative temperatures. We can extract the temperature evolution of the scattering rate (1/$\tau$) and plasma frequency ($\Omega_{p}$) from the fitting, as shown in Fig.~\ref{fit data}. As the temperature is increased, $\Omega_{p}$ decreases roughly, in accord well with the evolution of plasma edge observed in the reflective spectrum, and 1/$\tau$ increases monotonously. The change of $\Omega_{p}$ and 1/$\tau$ reveals a coherent optical conductivity, consistent well with the coherent interlayer behavior confirmed through transport-measurement in Ref.~\cite{Wang2018}.

A noteworthy result of this fitting is that a frequency-independent optical conductivity at various temperatures can be easily recognized in 300-800~cm$^{-1}$ range. The constant optical conductivity is analogous to the two-dimensional Dirac electron systems, such as the graphite and graphene~\cite{Kuzmenko2008,Mak2008,Ando2002}. However, for three-dimensional compounds, such unusual feature is a hallmark of the realized Dirac nodal-line dispersion near the Fermi level~\cite{Carbotte2017}, resembling the cases in NbAs$_{2}$~\cite{Shao2019} and ZrSiS~\cite{Schilling2017-2,Habe2018}.

\subsection{Theoretical discussion}

\begin{figure*}[tb]
  \centering
  \includegraphics[width=2\columnwidth]{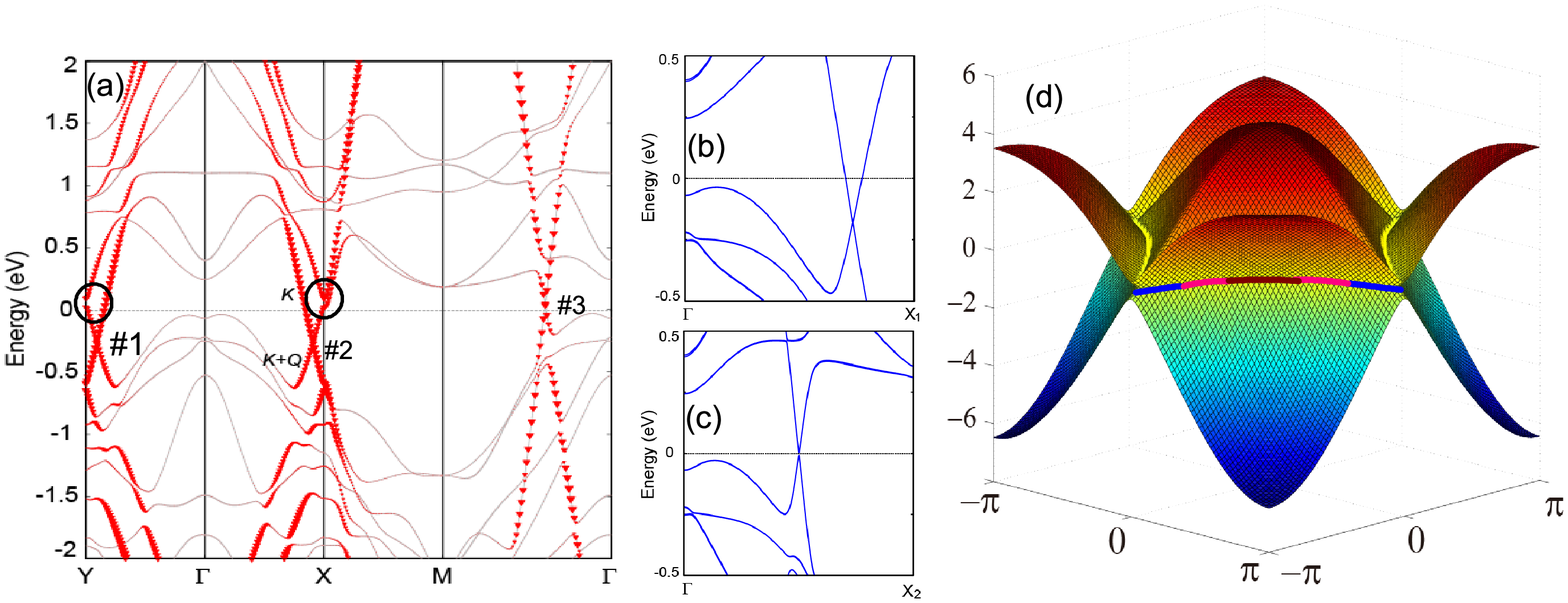}\\
  \caption{(a) Calculated band structure of YbMnSb$_{2}$ with spin-orbital coupling in the G-type antiferromagnetic order. The red color denotes the $p_{x/y}$ orbitals of Sb atom. The electronic structure from $\Gamma$ to the two representative points $X_{1}$ (b) and $X_{2}$ (c) along $M\sim X$. (d) The sketch shows the positions of Dirac bands change from $\Gamma\sim X$ to $\Gamma\sim M$. }
  \label{theoretical calculation}
\end{figure*}

In order to get more insight into the low frequency interband transitions, the first-principle methods have been used to calculate the bulk band structure. The crystal structure of YbMnSb$_{2}$ is depicted in Fig.~\ref{resistivity}(a). Our calculations are performed using density functional theory (DFT) as implemented in the Vienna ab initio simulation package (VASP) code \cite{Kresse1993,Kresse1996,Kresse1996-2}. The generalized-gradient approximation (GGA) for the exchange correlation functional is used. Throughout this work, the cutoff energy is set to be 500~eV for expanding the wave functions into plane-wave basis. In the calculation, the BZ is sampled in the k space within Monkhorst-Pack scheme\cite{Monkhorst1976}. On the basis of the equilibrium structure, the k mesh used is $10\times10\times4$. In the calculations we consider G-type antiferromagnetic and magnetic axis to be in plane \cite{Kealhofer2018}, and spin orbital coupling (SOC) is included. The experimental parameters are used in the calculations~\cite{Kealhofer2018}.

Figure~\ref{theoretical calculation}(a) shows the band structure with SOC in the G-type antiferromagnetic. Near the Fermi level, the electronic states are predominantly derived from the $p_{x/y}$ orbitals of Sb1 atom in the YbSb layer, and the $p$ orbitals of Sb2 are fully filled. Because of the crystal symmetry, the low-energy electronic band structures are almost the same along $\Gamma\sim X$ and $\Gamma\sim Y$ directions. The small volume crossing the Fermi surface agrees well with the low plasma frequency observed from our optical spectrum. Obviously, three Dirac-cone like dispersions can be recognized along $\Gamma\sim X$, $\Gamma\sim Y$, and $\Gamma\sim M$ (labeled with $\#1$, $\#2$ and $\#3$), and the gaps of which induced by SOC is about 10~meV. These three Dirac-cone like dispersions make it sufficient to consider the possibility of realizing the Dirac nodal-line state in YbMnSb$_{2}$.

To prove the nodal-circle electronic dispersion in YbMnSb$_{2}$, we randomly choose another two directions to calculate the band structures. Figure~\ref{theoretical calculation}(b) and (c) present the band structures along $\Gamma\sim X_{1}$ [$X_{1}$ = (1, 0.2, 0)] and $\Gamma\sim X_{2}$ [$X_{2}$ = (1, 0.5, 0)], from which two Dirac points can be easily confirmed. The different energies of these detected Dirac points indicate that the Dirac bands are shifted along the nodal loop. Together with the constant component identified in the real part of the optical conductivity, we can reasonably conclude that the Dirac nodal-line dispersion is realized in YbMnSb$_{2}$. The sketch of the three-dimensional band structures for the Dirac nodal-line is portrayed in Fig.~\ref{theoretical calculation}(d).

As SOC induces the gap on the nodal line, we can separate the quasi-two dimensional Dirac bands into three groups as shown in Fig.~\ref{theoretical calculation}(d): the Fermi level crosses the conduction bands (denoted by dark red solid line), the Fermi level cross the valence bands (denoted by blue solid lines), and the Fermi level within the SOC gap between conduction and valance bands (denoted by red solid lines).

In Ref.~\cite{Carbotte2017}, the characteristic optical response of ideal nodal-line semimetals is a constant $\sigma_{1}(\omega)$ at low frequency. However, for YbMnSb$_{2}$ we studied here,  in addition to the constant component, another Lorentz term is observed at low temperature. Generally, there are three possible scenarios for its origin:

(i) Charge-density-wave (CDW) or spin-density-wave (SDW). In LaAgSb$_{2}$~\cite{Chen2017} and iron based superconductors~\cite{Hu2008,Qazilbash2009,Charnukha2011}, a Lorentz-like response can be recognized before CDW or SDW transitions. For YbMnSb$_{2}$, even though it shows a long-range antiferromagnetic order~\cite{Wang2018} below 345 K, no CDW or SDW transition is reported in transport measurements~\cite{Kealhofer2018,Wang2018}. Moreover, as temperature increases, $\sigma_{1}(\omega)$ in the range of the Lorentz term is gradually filled instead of being suppressed, distinct from the optical feature caused by CDW or SDW transition~\cite{Chen2017,Hu2008}.

(ii) Indirect interband transitions. As mention in the book by Yu and Cardona \emph{et al.}, every indirect energy gap can give rise to two absorption edges at E$_{ig}$+E$_{p}$ and E$_{ig}$-E$_{p}$~\cite{Yu1999}, where E$_{ig}$ is the indirect energy gap and E$_{p}$ is the phonon energy. But these two absorption edges are not detected from the $\sigma_{\omega}$ spectra in YbMnSb$_{2}$ case. What is more, we notice that the direct transitions at X and Y points are about 100 meV in Fig.~\ref{theoretical calculation}(a)(denoted by the black circles), which are in accord with the frequency of Lorentz term. It is known that direct interband transitions have higher probability than indirect ones which require the assistance of phonons to conserve the momentum. Accordingly, the indirect interband transitions have little possibility in our case.

(iii) Multiple interband excitations. Besides the direct interband transitions mentioned above, the electronic excitations of nodal-line can also contribute to the Lorentz term. In Fig.~\ref{theoretical calculation}(d), we have separated the nodal-line Dirac bands into three parts, among which only the part denoted by red solid lines resembles the ideal nodal-line semimetals~\cite{Carbotte2017}, that can give rise to a constant optical component as we resolved in Fig.~\ref{fit}. The interband transitions of the blue and the dark red lines are supposed to have complex contributions to Lorentz term for their energy-shifted bands along the nodal loop. Therefore, the Lorentz term is ascribed to the direct interband transitions denoted by black circles in Fig.~\ref{theoretical calculation}(a), the blue and the dark red solid lines in nodal loop (in Fig.~\ref{theoretical calculation}(d)). Theoretically, both constant and Lorentz components should have a same onset energy because of the gap induced by SOC (10~meV). However, in our optical spectrum, it is hard to identify the particular onset energy at low temperature on account of the Drude term overlaps with the constant and Lorentz components, as a result of which the observed value (about 25meV) is roughly in agreement with the theoretical calculations. Overall, it is convinced that the multiple interband excitations contribute to the Lorentz term.

In the following, we analyze the stability of the nodal ring with or without SOC. When the SOC is ignored, the crossing bands are referred to as $K$ and $K+Q$ bands. Similar to iron-superconductors, the $K$ and $K+Q$ bands belong to the different eigenvalues of the glide plane symmetry, indicating that the nodal ring is robust. When the onsite SOC is included, the spin-nonflip term $<p_x,\sigma|H_{soc}|p_y,\sigma>$ only exists, and the spin-flip term $<p_x,\sigma|H_{soc}|p_y,\bar{\sigma}>$ is zero. Because $|p_y,\sigma>$ and $|p_x,\sigma>$ states have different eigenvalues of the glide plane symmetry, the term $<p_x,\sigma|H_{soc}|p_y,\sigma>$ also is zero. Therefore, the nodal ring is stability only in the $p_x$ and $p_y$ basis. However, if we consider p$_z$ orbital, an effective coupling between $|p_x,\sigma>$ and $|p_y,\bar{\sigma}>$ can be induced. The corresponding process can be summarized as
\begin{eqnarray}
|p_{x,\sigma}\rangle \stackrel{\lambda}{\longrightarrow}  |p_{z,\bar{\sigma}}\rangle  \stackrel{t}{\longrightarrow}  |p_{y,\bar{\sigma}}\rangle,
\end{eqnarray}
where $\sigma=\uparrow,\downarrow$ labels the spin, $\lambda$ is spin-orbital coupling and $t$ is hopping. The hybridization process is that $|p_{x,\sigma}\rangle$ couples strongly with $|p_{z,\bar{\sigma}}\rangle$ due to atomic SOC and $|p_{z,\bar{\sigma}}\rangle$ can hopping hybridize with $|p_{y,\bar{\sigma}}\rangle$. Finally, the effective coupling is proportional to $\lambda t$. In the Sb layer, the hoppings between p$_{x/y}$ and p$_z$ are prohibited due to glide plane symmetry, hence $t$ is an effective hopping by MnSb or Yb layers and fairly small. Therefore, if the effective coupling between $|p_x,\sigma>$ and $|p_y,\bar{\sigma}>$ is ignored, the nodal ring is also robust.

To our best knowledge, the experimentally confirmed nodal-line semimetals are found in those compounds whose weak SOC has little effect on the Dirac nodal-loop near the Fermi level~\cite{Schilling2017-2,Bian2016}, and the stronger SOC often triggers finite gaps at Dirac points in AMnSb$_{2}$/AMnBi$_{2}$ family~\cite{Qiu2018,Chaudhuri2017,Park2011,Farhan2014}. But in YbMnSb$_{2}$ case, the robust nodal ring is realized, offering the possibility of identifying more nodal-line behavior in pnictide Bi/Sb compounds. As the nodal-line is robust in YbMnSb$_{2}$ and it shares a similar crystal structure to iron based superconductors, more intriguing topological phenomena might be observed, such as superconductors with nodal lines. Owing to the existence of magnetic order alongside the nodal-line band structures, the realization of Weyl semimetals with broken time reversal symmetry, and even the pairing between magnetic order and Weyl fermions in YbMnSb$_{2}$ system might be detected~\cite{Armitage2018}.

\section{Conclusion}

To summarize, we have synthesized the single crystals of YbMnSb$_{2}$ and measured the detailed temperature and frequency dependence of the optical conductivity. In the real part of optical conductivity, a constant component and a Lorentz-like term have been resolved in the low frequency range at all measured temperatures. The constant component observed in $\sigma_{1}(\omega)$ indicates the possibility of topological nodal-line semimetal in YbMnSb$_{2}$. A range of theoretical calculations prove the existence of Dirac nodal-line dispersion in the Brillouin zone. In comparison with the calculated results, we conclude that the constant optical component is ascribed to the nodal-line interband transitions where Fermi level is within the SOC gap, and the Lorentz term is interpreted well with the multiple interband transitions. Together with theoretical analysis, we believe that YbMnSb$_{2}$ is a new kind of robust nodal-line semimetal.

\section{Acknowledgments}
We thank Hongtao Rong and Chunhong Li for useful discussions. This work was supported by NSFC (Projects No. 11774400 and No. 11404175) and MOST(973 Projects No. 2015CB921303, No. 2017YFA0302903 and No. 2015CB921102).

%
\bibliographystyle{apsrev}

\end{document}